\documentclass[twocolumn,showpacs,preprintnumbers,amsmath,amssymb]{revtex4}

\usepackage{graphicx}
\usepackage{dcolumn}
\usepackage{bm}

\newcommand*\samethanks[1][\value{footnote}]{\footnotemark[#1]}

\begin{document}

\title{Proton distribution radii of $^{12-19}$C illuminate features of neutron halos} 

\author {R. Kanungo$^{1}$, W. Horiuchi$^2$, G. Hagen$^{3,4}$\footnote{This manuscript has been authored by UT-Battelle, LLC under Contract
No. DE-AC05-00OR22725 with the U.S. Department of Energy. The United States Government retains and the publisher, by accepting the article for publication, acknowledges that the United States
Government retains a non-exclusive, paid-up, irrevocable, world-wide license to publish or reproduce the published form of
this manuscript, or allow others to do so, for United States Government purposes. The Department of Energy will provide
public access to these results of federally sponsored research in accordance with the DOE Public Access Plan
(http://energy.gov/downloads/doe-public-access-plan)}, 
G. R. Jansen$^{5,3}$\samethanks[1], P. Navratil$^{6}$, F. Ameil$^7$, J. Atkinson$^1$, Y. Ayyad$^8$, D. Cortina-Gil$^8$, I. Dillmann$^7$$\footnote{Present address : TRIUMF, Vancouver, Canada}$, A. Estrad\'e$^{1,7}$,  A. Evdokimov$^7$, F. Farinon$^7$, H. Geissel$^{7,9}$, G. Guastalla$^7$,  R. Janik$^{10}$,  M. Kimura$^2$, R. Kn\"obel$^7$, J. Kurcewicz$^7$, Yu. A. Litvinov$^7$, M. Marta$^7$, M. Mostazo$^8$, I. Mukha$^7$, C. Nociforo$^7$, H.J. Ong$^{11}$, S. Pietri$^7$, A. Prochazka$^7$,  C. Scheidenberger$^{7,9}$, B. Sitar$^{10}$, P. Strmen$^{10}$, Y. Suzuki$^{12,13}$, M. Takechi$^7$,  J. Tanaka$^{11}$,  I. Tanihata$^{11,14}$, S. Terashima$^{14}$, J. Vargas$^8$, H. Weick$^7$,  J. S. Winfield$^7$}

\affiliation{$^1$Astronomy and Physics Department, Saint Mary's University, Halifax, NS B3H 3C3, Canada}
\affiliation{$^2$Department of Physics, Hokkaido University, Sapporo 060-0810, Japan}
\affiliation{$^3$Physics Division, Oak Ridge National Laboratory, Oak Ridge, TN 37831, USA }
\affiliation{$^4$Department of Physics and Astronomy, University of Tennessee, Knoxville, TN 37996, USA}
\affiliation{$^5$National Center for Computational Sciences, Oak Ridge National Laboratory, Oak Ridge, TN 37831 USA}
\affiliation{$^6$TRIUMF, Vancouver, BC V6T 4A3, Canada}
\affiliation{$^7$GSI Helmholtzzentrum f\"ur Schwerionenforschung, D-64291 Darmstadt, Germany}
\affiliation{$^8$Universidad de Santiago de Compostela, E-15706 Santiago de Compostella, Spain}
\affiliation{$^{9}$Justus-Liebig University,  35392 Giessen, Germany}
\affiliation{$^{10}$Faculty of Mathematics and Physics, Comenius University, 84215 Bratislava, Slovakia}
\affiliation{$^{11}$RCNP, Osaka University, Mihogaoka, Ibaraki, Osaka 567 0047, Japan}
\affiliation{$^{12}$RIKEN Nishina Center, Wako, Saitama 351-0198, Japan}
\affiliation{$^{13}$Department of Physics, Niigata University, Niigata 950-2181, Japan}
\affiliation{$^{14}$School of Physics and Nuclear Energy Engineering and IRCNPC, Beihang University, Beijing 100191, China}

\date{\today}

\begin{abstract}

Proton radii of $^{12-19}$C densities derived from first accurate charge changing cross section
measurements at 900$A$ MeV with a carbon target are
reported. A thick neutron surface evolves from $\sim$ 0.5 fm
in $^{15}$C to $\sim$ 1 fm in $^{19}$C. The halo radius in $^{19}$C is
found to be 6.4$\pm$0.7 fm as large as $^{11}$Li. 
\emph{Ab initio} calculations based on chiral nucleon-nucleon and three-nucleon forces reproduce well the
radii.  

\end{abstract}

\maketitle

The existence of thick neutron skins
and halos \cite{TA85, HA87,TA13} in neutron-rich nuclei has brought a
dramatic change in our view of the nucleus.  These unexpected features
are exhibited through formation of neutron dominated nuclear surfaces
and hence large root mean square point matter radii ($R_m$).  The knowledge on how root mean square point proton distribution radii, 
henceforth in the article referred to
as proton radii ($R_p$),
evolve with neutron excess is still extremely limited. Proton radii are crucial
for deriving the neutron skin (surface) thickness and understanding the
spatial correlation between halo neutrons and its core-nucleus. Proton radii can
also  provide knowledge on 
shell structure evolution, as recently discussed for $^{52}$Ca \cite{RU16}.
The neutron skin may also be related to the symmetry energy (\emph{S}$_v$) and its density derivative at
saturation density (\emph{L}) defining the equation of state (EOS) of
asymmetric nuclear matter \cite{LA12}.  

Here we report the first precise determination of proton radii of neutron-rich isotopes $^{15-19}$C from the measurement of charge changing cross
sections that show rapidly growing thick neutron surfaces approaching the neutron drip line. The proton radii derived for $^{12-14}$C are in agreement with those
obtained from traditional methods such as electron scattering without any scaling factor. This clearly established the present technique as a valuable method to 
determine the proton radii of very neutron-rich isotopes. 
The measured radii  are in good agreement with those computed using \emph{ab initio} coupled-cluster theory based on chiral nucleon-nucleon and three-nucleon interactions \cite{HA15}.

The carbon isotopes draw interest because their $R_m$ show large
enhancements for $^{15,19}$C \cite{OZ01} and $^{22}$C \cite{HO06,TA10}.
This signals the presence of neutron halos. 
It is interesting to see how such structure evolution of
neutron-rich C isotopes affects their proton
distribution. 

Electron-nucleus scattering and measurement of muonic X-rays are
used to determine the charge radii ($R_c$) of stable
nuclei. The $R_c$ of $^{12}$C from e$^-$ scattering was found to be 2.478$\pm$0.009 fm \cite{OF91}
which is consistent with 2.472$\pm$0.015 fm from muonic X ray studies
\cite{SC82}. For $^{13}$C, the weighted average of two e$^-$
scattering measurements \cite{HE70,BE71} yields $R_c$ = 2.43$\pm$0.02
fm, while the muonic X-ray measurements \cite{SC82,BO85} find $R_c$
=2.463$\pm$0.004 fm. Results from e$^-$ scattering of $^{14}$C gives $R_c$
=2.56$\pm$0.05 fm \cite{KL73}, which is in agreement with 2.496$\pm$0.019 fm
from muonic X-ray measurements.

At present these techniques cannot be used for neutron-rich carbon
isotopes. A new approach, used in this work, is to measure the charge changing cross section ($\sigma_{cc}$) and 
derive the point proton radius ($R_p$) from it using the finite-range Glauber model. This
method has been employed in Refs.\cite{ES15,TE} and with zero-range calculations in Ref.\cite{YA11}. 
The effect of proton evaporation from neutron removal cross sections to states above the proton threshold is negligibly small for $^{12-19}$C since these nuclei are not in the vicinity of any proton unbound isotopes and the proton separation energies are fairly large.  At beam energies $\sim$900$A$ MeV, nuclear inelastic excitation cross section to states above the proton emission threshold is also negligibly small. The above effects become relevant for correction for nuclei at or neighbouring the proton drip-line. 

The first precise measurements of charge changing cross
sections and hence $R_p$ of neutron-rich isotopes $^{15-19}$C
as well as for $^{12-14}$C are reported here. The experiment was performed using the
fragment separator FRS \cite{GE92} at GSI, Darmstadt, Germany. The
carbon isotopes were produced through fragmentation of 1$A$ GeV
primary beams of $^{20}$Ne and $^{40}$Ar interacting with a 6.3
g/cm$^2$ thick Be target. The isotopes of interest were separated,
identified, and counted using event-by-event information of magnetic
rigidity (B$\rho$), time-of-flight (TOF), and energy-loss
($\Delta$E). A  multi-sampling ionization chamber (MUSIC) \cite{ST02}
provides the $Z$ identification from $\Delta$E. Figure 1 shows the experiment setup.
The first three (F1, F2 and F3), focal planes of the FRS
are dispersive while the final one, F4, is achromatic, where the
reaction target, C (4.01 g/cm$^2$) was placed. The energies of the isotopes at the
reaction target were $\sim$900$A$ MeV and are listed in Table 1. Plastic scintillator
detectors placed at the mid-plane F2 and before the reaction target at
F4 measured the time-of-flight of the incoming beam. The scintillator
before the target at F4 was used as the trigger of the data
acquisition system. Two position sensitive time projection chambers
(TPC)  \cite{HL98} were placed before the target at F4 which provided
beam tracking defining the beam profile on the target. In addition, TPCs
were also placed at F2. The position information in
combination with the central magnetic rigidity of the dipoles was used
to determine event by event B$\rho$ of the incident particle.

\begin{figure}
\includegraphics[width=9cm, height=4cm]{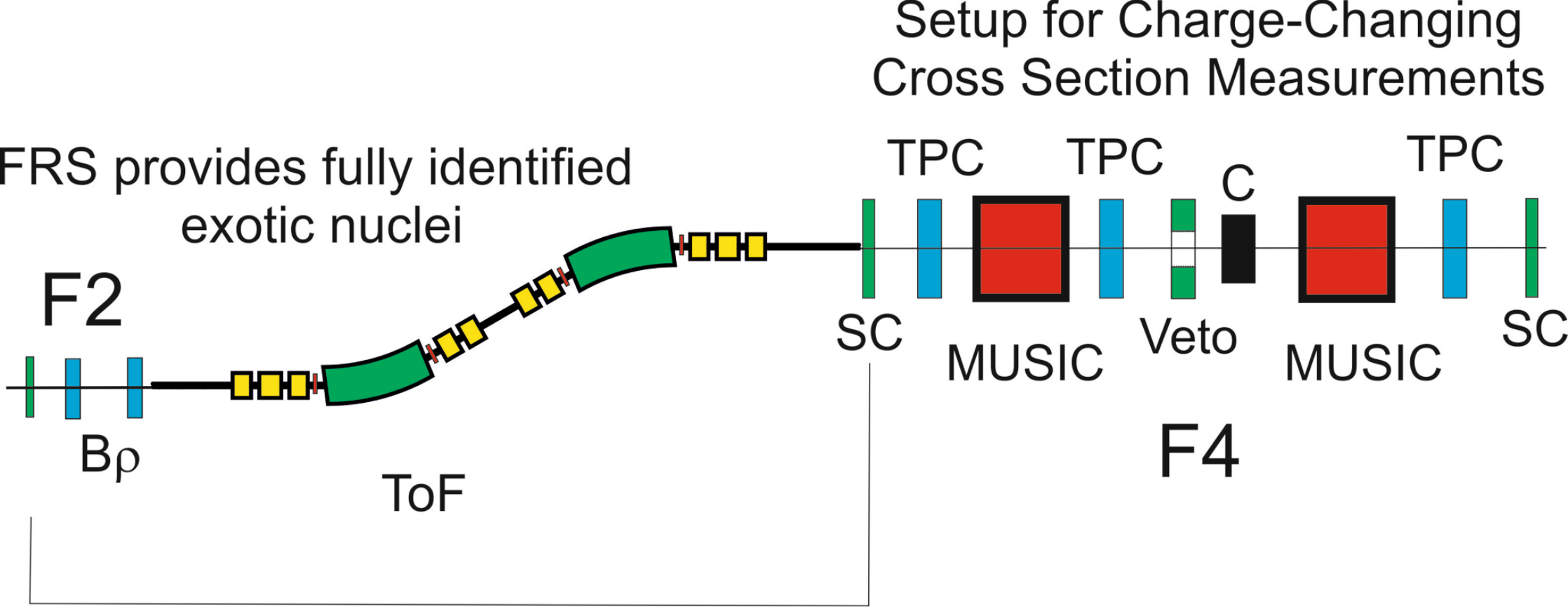}
\caption{\label{fig:epsart}  Schematic view of the experiment setup at the FRS with detector arrangement at the final focus F4. }
\end{figure}

The $\sigma_{cc}$ is measured using the transmission technique, where
the number ($N_{in}$) of incident nuclei $^AZ$, before the reaction
target is identified and counted. After the target, the nuclei with
the same charge $Z$ are identified and counted event by event
($N_{sameZ}$). The $\sigma_{cc}$ is obtained from a ratio of these
counts and is defined as $ \sigma_{cc} = t^{-1} {\mathrm {ln}} \big(
T_{t_{out}} / T_{t_{in}} \big) $ where $T=N_{sameZ}/N_{in}$, $t_{in}$ and
$t_{out}$ refer to measurements with and without the reaction target,
$t$ is the thickness of the target.

A restricted position and angle selection of the beam on target eliminated
spurious effects of losses due to large angle scattering out of the
detector acceptance. A veto scintillator was placed before the target
with a central hole of a size smaller than the target. This rejected
scattered events from upstream matter and multi-hit events where one
of the particles can miss hitting a MUSIC giving incorrect
reaction information. The particle identification condition of the
incident beam was defined in a way such that the contamination level
from $Z$=5 and 7 beam events relative to $Z$=6 is $\leq$10$^{-4}$.

After the reaction target the beam events with $Z$=6 were counted
using the second MUSIC. The energy-loss values of the TPC
and the plastic scintillator detectors placed after the target
provided additional information to ensure proper $Z$ identification
and counting. The resolution of the MUSIC for $Z$=$6$ was
$\Delta Z$ (in $\sigma$) = 0.12. The selection window covered $\sim \pm$4$\sigma$ of the
$Z$=6 particles.

With the desired isotope of C selected as the incoming beam, the
production of $Z$=7 events after the target is from charge exchange or
proton transfer reactions where one proton is added to the
nucleus. This cross section therefore does not involve reactions with
the protons in the C isotope and is hence subtracted to derive the
measured charge changing cross section. While this cross section is
generally very small ($<$1 mb), for the neutron-rich C isotopes it was
found to be a few mb \cite{TA16}.

\begin{figure}
\includegraphics[width=6cm, height=10cm]{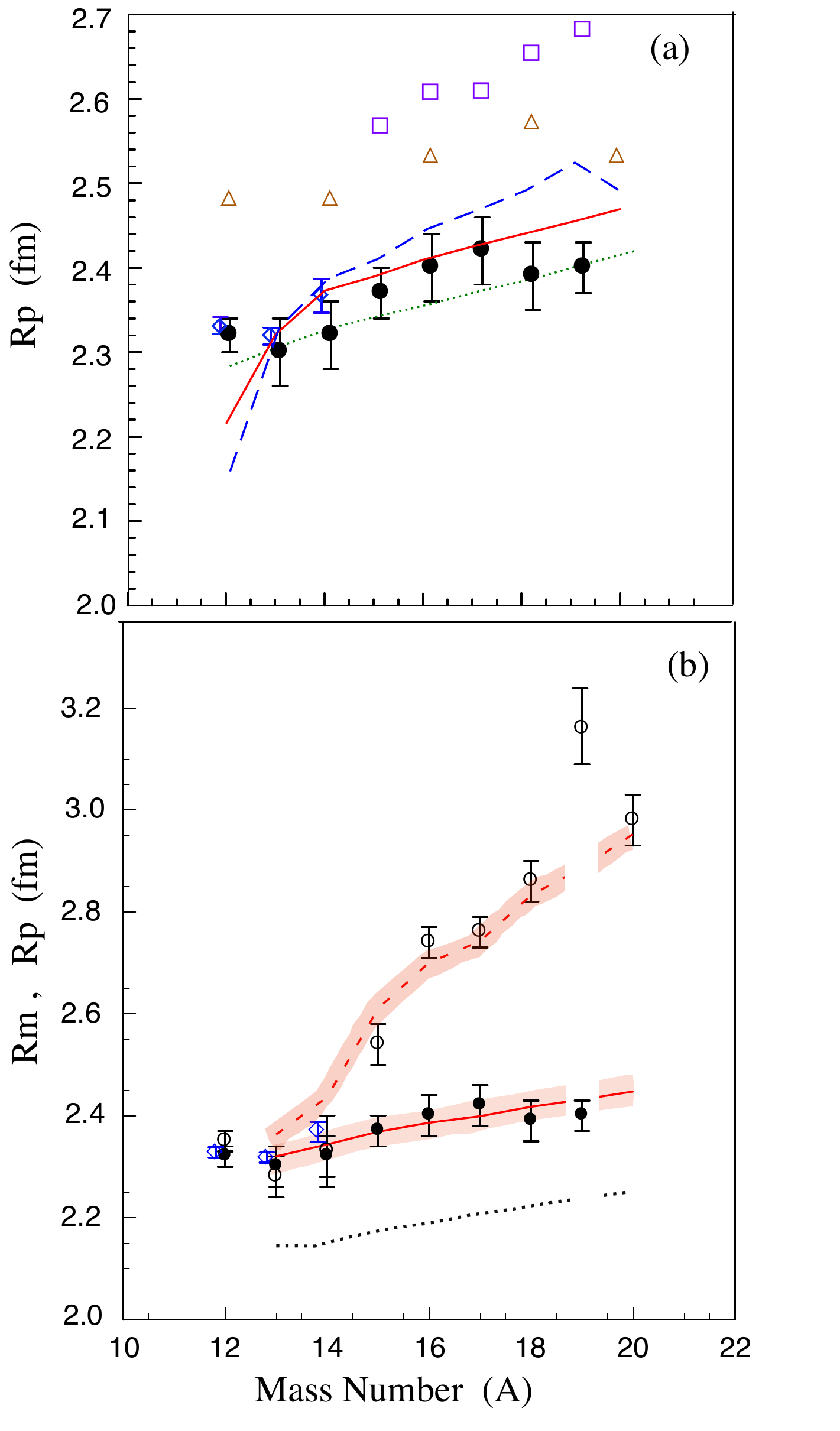}
\caption{\label{fig:epsart} $R_p$ extracted from
  $\sigma_{cc}$ for $^{12-19}$C (black filled circles). The blue open
  diamonds are $R_p$ from $e^-$-scattering and muonic X-rays
for $^{12-14}$C. (a) The relativistic mean field
  theory calculations with spherical/deformed potentials are shown by
  the red solid line/ blue dashed line. The green dotted line shows
  results of Hartree Fock calculations. The AMD results are shown by
  open triangles \cite{EN15} and open squares \cite{KI}. (b) The $R_m$
  from the present analysis of $\sigma_{I}$ \cite{OZ01} are shown by
  black open circles. The results of coupled cluster calculations are
  shown for $R_p$ as solid (red) and dotted (black) lines using the
  chiral NNLO$_{\rm sat}$ interaction and the nucleon-nucleon interaction NNLO$_{\rm opt}$  \cite{ekstrom2013}, respectively. 
  The matter radii with NNLO$_{\rm sat}$ is shown by the red dashed
  line. The shaded bands show the predicted uncertainty.}
\end{figure}

The measured values of $\sigma_{cc}$ are listed in Table 1. The cross section
increases for $^{15}$C which has a halo structure and continues
gradually increasing for $^{16,17}$C.  This is unlike Ref.\cite{YA11},
reporting the $\sigma_{cc}$ of $^{16}$C to be smaller than $^{15}$C.
With the halo-effect of $^{15}$C one would expect a small increase in the
proton radius as seen for example for $^{11}$Be \cite{NO09}.
A large increase in $\sigma_{cc}$ is not found for $^{19}$C although
it is a halo nucleus. This is because the effect of the center of mass motion of the halo on
the proton radius becomes smaller with larger mass number than in $^{11}$Be.
The $\sigma_{cc}$ for $^{12-20}$C reported in Ref.\cite{CH00} with large uncertainties
are systematically higher than those in \cite{WE90} for stable isotopes and not consistent with $R_p$
from e$^-$ scattering. 

The finite-range Glauber model \cite{SU} with harmonic oscillator
density is used for deriving the $R_p$ from the $\sigma_{cc}$. 
The $R_p$ are listed in Table 1. The charge radii of $^{12-14}$C known from e$^-$-
scattering and muonic X-ray measurements are used to
find the respective $R_p$ (blue diamonds in Fig.2 and R$_p^{(e^-,\mu)}$ in Table 1) following the formula in
Ref.\cite{NO09}. A good
agreement is seen with the $R_p$ found in this work. This lends strong
support to this technique of successfully extracting $R_p$ from
$\sigma_{cc}$. No scaling factor of $\sigma_{cc}$ was required for this agreement as was also 
seen in Ref.\cite{ES15}, unlike the discussion in Ref.\cite{YA11}. 

The matter radii of $^{12-19}$C shown in Fig.2b (open circles) are
derived in this work from a finite-range Glauber model \cite{HO07}  analysis 
of the interaction cross section $(\sigma_I)$ data from Ref.\cite{OZ01}. In
this analysis the $R_p$ are fixed to the values from Table 1
while the neutron radii are varied to reproduce the $(\sigma_I)$ data. The matter radius of $^{20}$C is shown from Ref.\cite{OZ01}. 
The combined information from proton radii and matter radii allow to fully characterize the halo features of $^{15,19}$C. 
 In a core plus neutron model following Ref.\cite{TA13}, the halo radius, R$_h$, of
$\sim$ 6.4$\pm$0.7 fm for $^{19}$C derived in this work shows the presence of a more
prominent halo in this nucleus compared to $\sim$ 4.2$\pm$0.5 fm for
$^{15}$C. The root mean square distance between the center of mass of the core and 
the halo neutron $R_{c-n}$ using the method in Ref.\cite{TA13}  for $^{15}$C is 7.2 $\pm$ 4.0 fm derived using $R_p$
and 4.15 $\pm$ 0.5 fm using $R_m$.  For $^{19}$C it is 4.1 $\pm$ 10 fm using $R_p$ and
6.6 $\pm$ 0.5 fm using $R_m$. The radius of the valence neutron in $^{19}$C was deduced to be 5.5$\pm$0.3 fm
from Coulomb dissociation \cite{NA99}.

The $R_p$ predicted by the relativistic mean field theory
\cite{SH15} using the NL3 parameters in a spherical potential is in 
overall agreement (Fig.~2a) with
the $R_p$ of $^{13-17}$C and slightly higher for $^{18,19}$C.
Those with a non-spherical potential  predict slightly higher radii for $^{14-19}$C.
The radii predicted in the framework of microscopic non-relativistic
Hartree-Fock method with a hybrid of Gogny and Skyrme effective
interactions is in agreement with the data of some isotopes \cite{SH15}. The radii
calculated in the Antisymmetrized Molecular Dynamics (AMD) framework greatly overpredict the data
in Fig.2a (open triangles  \cite{EN15}) and
 (open squares \cite{KI}).
 
We also perform coupled-cluster computations for the radii and
 compare with data. For the closed (sub-)shell nucleus $^{14}$C we use
 the coupled-cluster method with singles-and-doubles excitations
 \cite{bartlett2007} to compute the expectation value of the intrinsic
 point-proton and neutron radii. To access the open-shell nuclei
 $^{13,15}$C we use particle-removed/attached equation-of-motion
 coupled-cluster method \cite{gour2006,hagen2014}, while for
 $^{16-19}$C we employ the recently developed coupled-cluster
 effective interaction (CCEI) method in the $sd$ shell
 \cite{jansen2014,jansen2015}.  To compute the intrinsic radii of
 $^{16-19}$C within CCEI we follow the scheme outlined in Ref.\cite{jansen2014},
 and include the core and one-body parts of the valence-space radius
 operator, while we neglect the two-body part. We solve our
 coupled-cluster equations using a Hartree-Fock basis built from a
 harmonic-oscillator basis consisting of fifteen major oscillator
 shells ($N_{\rm max} = 2n+l = 14$) with the additional energy cut
 $E_{3\text{max}}=N_1+N_2+N_3 \leq 16$ for the three-nucleon
 interaction. Here $N_i = 2n_i + l_i$ refers to the major oscillator
 shell of the $i^{th}$ particle. For the computed radii we estimate an
 uncertainty of $\pm$0.04 fm coming from the model-space and
 coupled-cluster method.

 We perform our coupled-cluster calculations using various
 state-of-the-art chiral interactions. First, we focus on NNLO$_{{\rm
     sat}}$ which was obtained using a novel optimization strategy
 that simultaneously optimized the low-energy constants in the nucleon-nucleon (NN) and
 three-nucleon (3NF) sector at next-to-next-to leading order (NNLO)
 including data on charge radii and binding energies of selected
 nuclei up to $^{25}$O in the fit \cite{EK15}. NNLO$_{{\rm sat}}$ was
 recently successfully applied to compute radii of $^{48}$Ca
 \cite{HA15}, and is for the first time employed for $^{13-19}$C in
 this work.  Fig. 2b shows a comparison between data and
 coupled-cluster computations using NNLO$_{\rm sat}$ for the
 point-proton and matter radii of $^{13-19}$C. In addition to
 NNLO$_{{\rm sat}}$ we also compare data with the chiral interaction
 NNLO$_{{\rm opt}}$ \cite{ekstrom2013} which does not include 3NFs. We
 observe that results with the NNLO$_{{\rm sat}}$ gives overall good agreement with
 data, while those with the NNLO$_{{\rm opt}}$ interaction significantly underestimate the
 radii. It is seen that the effects of simultaneously optimizing the
 low-energy coupling constants in the NN and three-nucleon sector, the
 inclusion of binding energies and radii of selected nuclei with
 $A\leq 25$ in the objecive function, and the inclusion of 3NFs with
 non-local regulators are indeed very significant and crucial for
 reproducing the measured proton radii. The results for the matter
 radii (Fig.2b dashed red curve) using NNLO$_{\rm sat}$ are also in
 good agreement with the data.  Since $^{19}$C is a weakly bound
 nucleus and the coupling to the particle continuum is not included in
 these calculations the CCEI result with NNLO$_{\rm sat}$ leads to
 $^{19}$C being unbound.  For this reason the radius is not defined
 and not shown in Fig. 2.

\begin{figure}
\includegraphics[width=8cm, height=4cm]{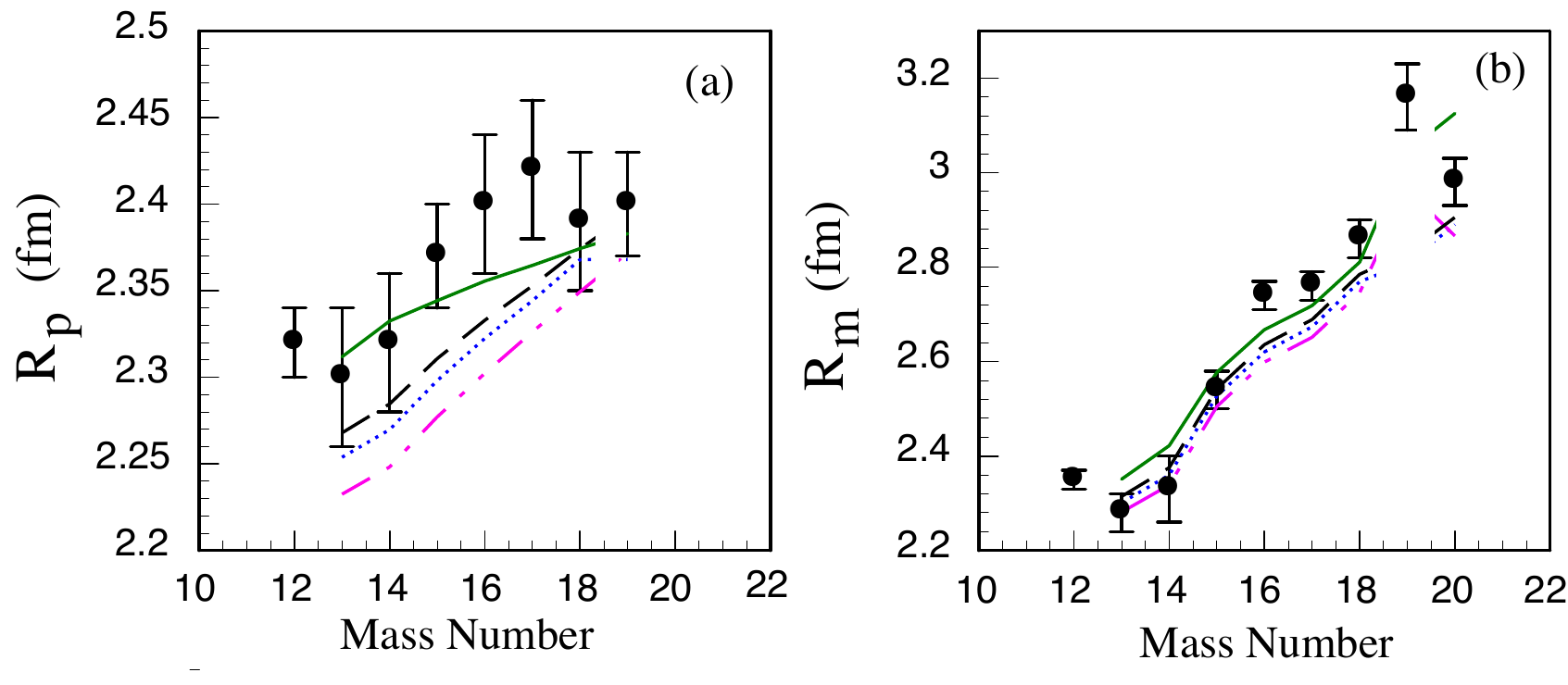}
\caption{\label{fig:epsart}  The comparison of data with coupled cluster predictions for (a) proton radii (b) matter radii.
The chiral interactions \cite{HE11} are EM1 (dotted blue curve), EM3 (dashed double-dotted pink curve),
EM4 (dashed black curve) and EM5 (solid green curve). }
\end{figure}

With the successful description of the radii in the coupled cluster
framework using the NNLO$_{\rm sat}$ interaction, we now investigate
how the data compares to a set of other chiral interactions. These
interactions are adopted from Ref.\cite{HE11} and include NN and
3NFs. The NN interactions are based on a similarity renormalization
group transformation \cite{bogner2007} of the chiral interaction at N$^3$LO
from Ref.~\cite{entem2003} and with a non-local 3NF at N$^2$LO. In
contrast to NNLO$_{\rm sat}$ the low-energy coupling constants were
determined from a fit to scattering data, binding energies and radii of
nuclei with $A\leq 4$. The different forces used have different NN (3NF) 
cutoffs, namely EM1 = 2.0 fm$^{-1}$ (2.0 fm$^{-1}$), EM3 = 1.8
fm$^{-1}$  (2.0 fm$^{-1}$), EM4= 2.2 fm$^{-1}$ (2.0 fm$^{-1}$), EM5 =
2.8 fm$^{-1}$ (2.0 fm$^{-1}$).  Fig. 3a and Fig. 3b compare the data of
proton radii and matter radii, respectively, with predictions using
these interactions that are shown as EM1 (dotted blue curve), EM3
(dashed double-dotted pink curve), EM4 (dashed black curve) and EM5
(solid green curve). It is seen that the agreement with the data over
the different isotopes with the NNLO$_{\rm sat}$ is much better than
any of the "EM" interactions. The $R_p$ for $^{13-17}$C are not
reproduced by the EM3 interaction. The EM1 and EM4 interactions do
not reproduce the measured $R_p$ for $^{15-17}$C. The interactions
with a lower NN cutoff seems to predict smaller radii values.  The
agreement of the predictions with the matter radii (Fig. 3b) is better
using the EM5 interaction than the other "EM" interactions, though
once again the NNLO$_{\rm sat}$ predictions seem to be in much better
agreement overall. This suggests that the NNLO$_{\rm sat}$ interaction
has a better predictive capability for bulk properties of nuclei such
as nuclear radii.

\begin{figure}
\includegraphics[width=8cm, height=4cm]{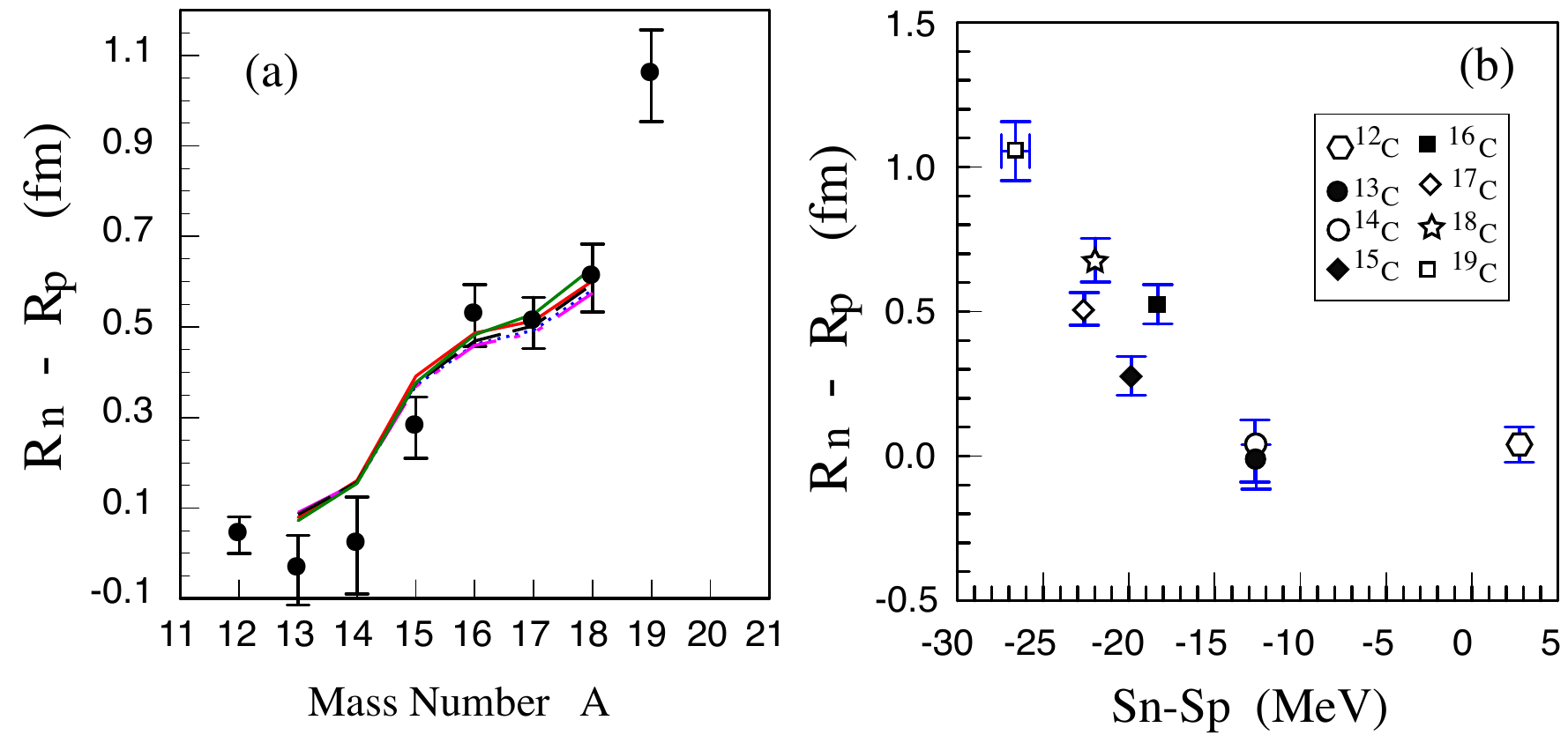}
\caption{\label{fig:epsart}  (a) The measured neutron skin thickness for $^{12-19}$C compared to predictions using the different
interactions, NNLO$_{\rm sat}$ (red solid curve), EM1 (dotted blue curve), EM3 (dashed double-dotted pink curve),
EM4 (dashed black curve) and EM5 (solid green curve).(b) The measured neutron skin thickness variation with $S_n - S_p$ for $^{12-19}$C.}
\end{figure}

The neutron skin thickness  defined as the difference of point neutron radius ($R_n$) and 
$R_p$ is shown in Fig.4a. The red
curve shows the coupled cluster calculations with the NNLO$_{\rm sat}$
interaction to be in good agreement with the data. The predictions with the
other interactions are all very similar to each other and to that with the
NNLO$_{\rm sat}$ interaction. We see a very thick
neutron skin developing with increasing neutron-proton asymmetry. In
Fig. 4b the relationship  of the neutron skin thickness as a function of the
difference between the one-neutron separation
energy ($S_n$) and one-proton separation energy ($S_p$) is shown.  The strong correlation observed in Fig. 4b points to a thick neutron skin (surface) being associated with the large Fermi-level difference of neutrons and protons, that occurs as nuclei become highly neutron-rich.

In summary, the first accurate determination of $R_p$ of $^{12-19}$C  is accomplished from 
charge changing cross section ($\sigma_{cc}$) measurements with a carbon target at 900$A$
MeV. The Glauber model successfully relates the $\sigma_{cc}$ to $R_p$ which is seen from their agreement with
radii from electron scattering. The  radii are in overall good agreement with 
coupled-cluster computations using the chiral interaction NNLO$_{\rm sat}$.

\begin{table}
\caption{\label{tab:table1} Secondary beam energies, measured $\sigma_{cc}$ and the root mean square proton and matter radii derived from the 
data for the carbon isotopes.}
\begin{ruledtabular}
\begin{tabular}{llllll}
Isotope & E/A &$\sigma_{cc}^{ex}$&R$_p^{ex}$&R$_p^{(e^-,\mu)}$ &R$_m^{ex}$\\
&(MeV)&(mb)&(fm)&(fm)&(fm)\\
\hline
$^{12}$C& 937  & 733(7)&2.32(2) &2.33(1) &2.35(2) \\
$^{13}$C& 828 & 726(7)&2.30(4)&2.32(1) &2.28(4)\\
$^{14}$C& 900 & 731(7)&2.32(4)& 2.37(2)&2.33(7)  \\
$^{15}$C& 907 & 743(7)&2.37(3)& & 2.54(4) \\
$^{16}$C& 907 & 748(7)&2.40(4)& &2.74(3) \\
$^{17}$C& 979 & 754(7)&2.42(4)&&2.76(3)\\
$^{18}$C& 895 & 747(7)&2.39(4)& &2.86(4)\\
$^{19}$C& 895 & 749(9)&2.40(3)& &3.16(7)\\
\end{tabular}
\end{ruledtabular}
\end{table}

The authors are thankful for the support of the GSI accelerator staff
and the FRS technical staff for an efficient running of the
experiment. The support from NSERC, Canada for this work is gratefully
acknowledged. R. Kanungo thankfully acknowledges the HIC-for-FAIR
program and JLU-Giessen for supporting part of the research stay. The
support of the PR China government and Beihang university under the
Thousand Talent program is gratefully acknowledged.  The experiment is
partly supported by the grant-in-aid program of the Japanese
government under the contract number 23224008. This work was
supported by the Office of Nuclear Physics, U.S. Department of Energy
(Oak Ridge National Laboratory), DE-SC0008499 (NUCLEI SciDAC
collaboration), NERRSC Grant No.\ 491045-2011, and the Field Work
Proposal ERKBP57 at Oak Ridge National Laboratory.  Computer time was
provided by the Innovative and Novel Computational Impact on Theory
and Experiment (INCITE) program.  TRIUMF receives funding via a
contribution through the National Research Council Canada. This
research used resources of the Oak Ridge Leadership Computing Facility
located in the Oak Ridge National Laboratory, which is supported by
the Office of Science of the Department of Energy under Contract No.
DE-AC05-00OR22725, and used computational resources of the National
Center for Computational Sciences and the National Institute for
Computational Sciences.


\end{document}